# TDGL Equations for Two-Band Superconductor

V.L.OSTROVSKI

ORT Braude college, Karmiel, 21982, Israel

## Abstract

Green's function techniques for studying nonequilibrium processes in two-band superconductor are discussed. Perturbation expansions and Green's function equations are developed. A time dependent modification of the Ginsburg-Landau equation is given.

## Introduction.

Time dependent Ginzburg-Landau (TDGL) equation has proven to be very successful in the theory of superconductivity [1-11]. The usual classification of superconductors characterizes materials by one-component order parameter. However the properties of some compounds (e.g. $MgB_2, Nb_2Se, V_3Si$) are described by two-band model [12] with two-component order parameter. The two-gap GL model (the two gaps arise from the existence of two bands) has been proposed [13, 14] at the dawn of superconductivity theory. The interest to the two-gap model has recently been renewed after the discovery the superconductivity in magnesium diboride [15-42]. The studies below are based on a set of Ginzburg-Landau equations for the order parameters of a two-band system. We focus our attention mainly on the overlapping of energy bands on the Fermi surface and on anisotropy due to different dimensions of the energy bands under study.

E-mail: vostrovski@gmail.com

# 1. Green's functions in the Gor'kov representation.

The two-band pairing Hamiltonian in the BCS theory in the absence of external field is given by

$$\hat{H} = \hat{H}_0 + \hat{H}_{int} \qquad (1.1)$$

Where

$$\hat{H}_0 = \int d^3\vec{r} \sum_{n\sigma} \hat{\psi}^+_{n\sigma}(\vec{r}) \hat{\varepsilon}_n(\hat{\vec{p}}) \hat{\psi}_{n\sigma}(\vec{r}) \qquad (1.2)$$

is the Hamiltonian of noninteracting quasiparticles, $\hat{\varepsilon}_n(\hat{\vec{p}}) = \zeta_n + \frac{\hat{\vec{p}}^2}{2m_n} - \mu$, $\mu$ is the chemical potential, $n=1,2$ are the band indexes, $\varepsilon_n(\vec{p}) = \zeta_n + \frac{\vec{p}^2}{2m_n} - \mu$ is the dispersion relation of quasiparticles in the n-th band, $m_n$ is the effective mass, and $\hat{\psi}^+_{n\sigma}, \hat{\psi}_{n\sigma}$ are the creation and annihilation field operators in the Heisenberg representation for quasiparticles in band $n$ with spin $\sigma = \uparrow or \downarrow$.

$$\hat{H}_{int} = -\sum_{nm} V_{nm} \int d^3\vec{r} \sum_{\sigma\sigma'} \hat{\psi}^+_{n\sigma}(\vec{r}) \hat{\psi}^+_{n\sigma'}(\vec{r}) \hat{\psi}_{m\sigma'}(\vec{r}) \hat{\psi}_{m\sigma}(\vec{r})(1-\delta_{\sigma\sigma'}) =$$

$$= -\sum_{nm} V_{nm} \int d^3\vec{r} [\ \hat{\psi}^+_{n\uparrow}(\vec{r}) \hat{\psi}^+_{n\downarrow}(\vec{r}) \hat{\psi}_{m\downarrow}(\vec{r}) \hat{\psi}_{m\uparrow}(\vec{r}) + \qquad (1.3)$$

$$+ \hat{\psi}^+_{n\downarrow}(\vec{r}) \hat{\psi}^+_{n\uparrow}(\vec{r}) \hat{\psi}_{m\uparrow}(\vec{r}) \hat{\psi}_{m\downarrow}(\vec{r})\ ]$$

is the Hamiltonian of interaction, $V_{11}, V_{22}$ are the constants of intraband interaction, $V_{12}, V_{21}$ are the constants of interband interaction $(V_{nm} = V_{mn})$. The model assumes the formation of Cooper pairs in each band and the tunneling of these pairs as a whole from one band into another. The equation of motion for Heisenberg operators $\hat{\psi}_n(\vec{r},t), \hat{\psi}_n^+(\vec{r},t)$ have the form.

The formulas (1.11) are equivalent to

$$i\frac{\partial \hat{\psi}_{n\sigma}(\vec{r},t)}{\partial t} = \hat{\varepsilon}_n \hat{\psi}_{n\sigma}(\vec{r},t) - \sum_{m\sigma'} V_{nm} \hat{\psi}_{m\sigma'}^+(\vec{r},t)\hat{\psi}_{m\sigma'}(\vec{r},t)\hat{\psi}_{n\sigma}(\vec{r},t)(1-\delta_{\sigma\sigma'}),$$

$$i\frac{\partial \hat{\psi}_{n\sigma}^+(\vec{r},t)}{\partial t} = -\hat{\varepsilon}_n \hat{\psi}_{n\sigma'}^+(\vec{r},t) + \sum_{m\sigma'} V_{nm} \hat{\psi}_{n\sigma}^+(\vec{r},t)\hat{\psi}_{m\sigma'}^+(\vec{r},t)\hat{\psi}_{m\sigma'}(\vec{r},t)(1-\delta_{\sigma\sigma'}), \qquad (1.4)$$

Inserting $\hat{\varepsilon}_n = \frac{\hat{\vec{p}}^2}{2m_n} - \mu + \zeta_n$ in the (1.4) we obtain

$$i\frac{\partial \hat{\psi}_{n\sigma}(\vec{r},t)}{\partial t} = \left(\frac{\hat{\vec{p}}^2}{2m_n} - \mu + \zeta_n\right)\hat{\psi}_{n\sigma}(\vec{r},t) - \sum_{m\sigma'} V_{nm} \hat{\psi}_{m\sigma'}^+(\vec{r},t)\hat{\psi}_{m\sigma'}(\vec{r},t)\hat{\psi}_{n\sigma}(\vec{r},t)(1-\delta_{\sigma\sigma'}),$$

$$i\frac{\partial \hat{\psi}_{n\sigma}^+(\vec{r},t)}{\partial t} = -\left(\frac{\hat{\vec{p}}^2}{2m_n} - \mu + \zeta_n\right)\hat{\psi}_{n\sigma'}^+(\vec{r},t) + \sum_{m\sigma'} V_{nm} \hat{\psi}_{n\sigma}^+(\vec{r},t)\hat{\psi}_{m\sigma'}^+(\vec{r},t)\hat{\psi}_{m\sigma'}(\vec{r},t)(1-\delta_{\sigma\sigma'}) \quad (1.5)$$

The Green's function is defined by

$$G_{n\sigma,m\sigma'}(\vec{r}_1,t_1;\vec{r}_2,t_2) = -i\left\langle T_t\left(\hat{\psi}_{n\sigma}(\vec{r}_1,t_1)\hat{\psi}_{m\sigma'}^+(\vec{r}_2,t_2)\right)\right\rangle_{N,N} \qquad (1.6)$$

where $\langle...\rangle_{N,N'}$ denotes the statistical averaging of matrix element between states with N and N' particles.

For systems having no net spin polarization

$$G_{n\sigma,m\sigma'}(\vec{r}_1,t_1;\vec{r}_2,t_2) = \delta_{\sigma\sigma'}G_{nm}(\vec{r}_1,t_1;\vec{r}_2,t_2)$$

Then

$$i\frac{\partial}{\partial t_1}G_{n\sigma,m\sigma'}(\vec{r}_1,t_1;\vec{r}_2,t_2) = -i\left\langle T_t\left(i\frac{\partial}{\partial t_1}\hat{\psi}_{n\sigma}(\vec{r}_1,t_1)\hat{\psi}^+_{m\sigma'}(\vec{r}_2,t_2)\right)\right\rangle_{N,N} =$$

$$= -i\left\langle \left[T_t\left((\frac{\hat{\vec{p}}_1^2}{2m_n} - \mu + \zeta_n)\hat{\psi}_{n\sigma}(\vec{r}_1,t_1) - \sum_{l\sigma''}V_{nl}\hat{\psi}^+_{l\sigma''}(\vec{r}_1,t_1)\hat{\psi}_{l\sigma''}(\vec{r}_1,t_1)\hat{\psi}_{n\sigma}(\vec{r}_1,t_1)(1-\delta_{\sigma\sigma''})\right)\hat{\psi}^+_{m\sigma'}(\vec{r}_2,t_2)\right]\right\rangle_{N,N} =$$

$$= (\frac{\hat{\vec{p}}_1^2}{2m_n} - \mu + \zeta_n)(-i)\left\langle T_t\left(\hat{\psi}_{n\sigma}(\vec{r}_2,t_2)\hat{\psi}^+_{m\sigma'}(\vec{r}_1,t_1)\right)\right\rangle_{N,N} -$$

$$-\sum_{l\sigma''}V_{nl}(-i)\left\langle T_t\left(\hat{\psi}^+_{l\sigma''}(\vec{r}_1,t_1)\hat{\psi}_{l\sigma''}(\vec{r}_1,t_1)\hat{\psi}_{n\sigma}(\vec{r}_1,t_1)\hat{\psi}^+_{m\sigma'}(\vec{r}_2,t_2)(1-\delta_{\sigma\sigma''})\right)\right\rangle_{N,N} -$$

$$-\delta(\vec{r}_1 - \vec{r}_2)\delta(t_1 - t_2)\delta_{\sigma\sigma'} =$$

$$= (\frac{\hat{\vec{p}}_1^2}{2m_n} - \mu + \zeta_n)G_{n\sigma,m\sigma'}(\vec{r}_1,t_1;\vec{r}_2,t_2) - i\sum_{l\sigma''}V_{nl}(-i)\left\langle T_t\left(\hat{\psi}_{l\sigma''}(\vec{r}_1,t_1)\hat{\psi}_{n\sigma}(\vec{r}_1,t_1)\right)\right\rangle_{N,N+2} \times$$

$$\times(-i)\left\langle T_t\left(\hat{\psi}^+_{l\sigma''}(\vec{r}_1,t_1)\hat{\psi}^+_{m\sigma'}(\vec{r}_2,t_2)\right)\right\rangle_{N+2,N}(1-\delta_{\sigma\sigma''}) - \delta(\vec{r}_1-\vec{r}_2)\delta(t_1-t_2)\delta_{\sigma\sigma'} =$$

$$= (\frac{\hat{\vec{p}}_1^2}{2m_n} - \mu + \zeta_n)G_{n\sigma,m\sigma'}(\vec{r}_1,t_1;\vec{r}_2,t_2) - i\sum_{l\sigma''}V_{nl}F_{l\sigma'',n\sigma}(\vec{r}_1,t_1;\vec{r}_1,t_1)F^+_{l\sigma'',m\sigma'}(\vec{r}_1,t_1;\vec{r}_2,t_2)(1-\delta_{\sigma\sigma''}) -$$

$$-\delta(\vec{r}_1-\vec{r}_2)\delta(t_1-t_2)\delta_{\sigma\sigma'}$$

where $F^+_{n\sigma,m\sigma'}(\vec{r}_1,t_1;\vec{r}_2,t_2) = -i\left\langle T_t\left(\hat{\psi}^+_{n\sigma}(\vec{r}_1,t_1)\hat{\psi}^+_{m\sigma'}(\vec{r}_2,t_2)\right)\right\rangle_{N+2,N}$

and $F_{n\sigma,m\sigma'}(\vec{r}_1,t_1;\vec{r}_2,t_2) = -i\left\langle T_t\left(\hat{\psi}_{n\sigma}(\vec{r}_1,t_1)\hat{\psi}_{m\sigma'}(\vec{r}_2,t_2)\right)\right\rangle_{N,N+2}$

are the anomalous Green's functions which satisfy the relations

$$F_{n\sigma,m\sigma'}(\vec{r}_1,t_1;\vec{r}_2,t_2) = g_{\sigma\sigma'}F_{nm}(\vec{r}_1,t_1;\vec{r}_2,t_2)$$

$$F^+_{n\sigma,m\sigma'}(\vec{r}_1,t_1;\vec{r}_2,t_2) = g_{\sigma\sigma'}F^+_{nm}(\vec{r}_1,t_1;\vec{r}_2,t_2)$$

$$g_{\sigma\sigma'} = \begin{pmatrix} 0 & 1 \\ -1 & 0 \end{pmatrix}$$

For the function $G_{nm}(\vec{r}_1,t_1;\vec{r}_2,t_2)$ we obtain the following equation

$$\left[i\frac{\partial}{\partial t_1} - \frac{\hat{\vec{p}}_1^2}{2m_n} + \mu - \zeta_n\right]G_{nm}(\vec{r}_1,t_1;\vec{r}_2,t_2) + i\sum_l V_{nl}F_{ln}(\vec{r}_1,t_1;\vec{r}_1,t_1)F^+_{lm}(\vec{r}_1,t_1;\vec{r}_2,t_2) = \qquad (1.7)$$
$$= \delta(\vec{r}_1 - \vec{r}_2)\delta(t_1 - t_2)$$

The function $F^+_{n\sigma,m\sigma'}(\vec{r}_1,t_1;\vec{r}_2,t_2)$ satisfies the equation

$$i\frac{\partial}{\partial t_1}F^+_{n\sigma,m\sigma'}(\vec{r}_1,t_1;\vec{r}_2,t_2) = -i\left\langle T_t\left(i\frac{\partial}{\partial t_1}\hat{\psi}^+_{n\sigma}(\vec{r}_1,t_1)\hat{\psi}^+_{m\sigma'}(\vec{r}_2,t_2)\right)\right\rangle_{N+2,N} =$$

$$= -i\left\langle T_t\left(\left[-\left(\frac{\hat{\vec{p}}_1^2}{2m_n}-\mu+\zeta_n\right)\hat{\psi}^+_{n\sigma}(\vec{r}_1,t_1) + \sum_{l\sigma''}V_{nl}\hat{\psi}^+_{n\sigma}(\vec{r}_1,t_1)\hat{\psi}^+_{l\sigma''}(\vec{r}_1,t_1)\hat{\psi}_{l\sigma''}(\vec{r}_1,t_1)(1-\delta_{\sigma\sigma''})\right]\hat{\psi}^+_{m\sigma'}(\vec{r}_2,t_2)\right)\right\rangle_{N+2,N} =$$

$$= -(\frac{\hat{\vec{p}}_1^2}{2m_n}-\mu+\zeta_n)\left[-i\left\langle T_t\left(\hat{\psi}^+_{n\sigma}(\vec{r}_1,t_1)\hat{\psi}^+_{m\sigma'}(\vec{r}_2,t_2)\right)\right\rangle_{N+2,N}\right] +$$

$$+\sum_{l\sigma''}V_{nl}\left[-i\left\langle T_t\left(\hat{\psi}^+_{n\sigma}(\vec{r}_1,t_1)\hat{\psi}^+_{l\sigma''}(\vec{r}_1,t_1)\hat{\psi}_{l\sigma''}(\vec{r}_1,t_1)\hat{\psi}^+_{m\sigma'}(\vec{r}_2,t_2)\right)\right\rangle_{N+2,N}(1-\delta_{\sigma\sigma''})\right] =$$

$$= -(\frac{\hat{\vec{p}}_1^2}{2m_n}-\mu+\zeta_n)F^+_{n\sigma,m\sigma'}(\vec{r}_1,t_1;\vec{r}_2,t_2) + i\sum_{l\sigma''}V_{nl}(-i)\left\langle T_t\left(\hat{\psi}^+_{n\sigma}(\vec{r}_1,t_1)\right)\hat{\psi}^+_{l\sigma''}(\vec{r}_1,t_1)\right\rangle_{N+2,N} \times$$

$$\times(-i)\left\langle T_t\left(\hat{\psi}_{l\sigma''}(\vec{r}_1,t_1)\hat{\psi}^+_{m\sigma'}(\vec{r}_2,t_2)\right)\right\rangle_{N,N}(1-\delta_{\sigma\sigma''}) = -(\frac{\hat{\vec{p}}_1^2}{2m_n}-\mu+\zeta_n)F^+_{n\sigma,m\sigma'}(\vec{r}_1,t_1;\vec{r}_2,t_2) +$$

$$+i\sum_{l\sigma''}V_{nl}F^+_{n\sigma,l\sigma''}(\vec{r}_1,t_1;\vec{r}_1,t_1)G_{l\sigma'',m\sigma'}(\vec{r}_1,t_1;\vec{r}_2,t_2)(1-\delta_{\sigma\sigma''})$$

or

$$\left[i\frac{\partial}{\partial t_1}+\frac{\hat{\vec{p}}_1^2}{2m_n}-\mu+\zeta_n\right]F^+_{nm}(\vec{r}_1,t_1;\vec{r}_2,t_2)-i\sum_l V_{nl}F^+_{nl}(\vec{r}_1,t_1;\vec{r}_1,t_1)G_{lm}(\vec{r}_1,t_1;\vec{r}_2,t_2)=0 \quad (1.8)$$

Since the nondiagonal functions are supposed neglected
$G_{12}, G_{21} \ll G_{11}, G_{22}$,
$F_{12}, F_{21} \ll F_{11}, F_{22}$

the system (1.7),(1.8) takes the form

$$\left[i\frac{\partial}{\partial t_1}-\frac{\hat{\vec{p}}_1^2}{2m_n}+\mu-\zeta_n\right]G_{nn}(\vec{r}_1,t_1;\vec{r}_2,t_2)+i\sum_l V_{nl}F_{ll}(\vec{r}_1,t_1;\vec{r}_1,t_1)F^+_{nn}(\vec{r}_1,t_1;\vec{r}_2,t_2)=$$
$$=\delta(\vec{r}_1-\vec{r}_2)\delta(t_1-t_2) \quad (1.9)$$

$$\left[i\frac{\partial}{\partial t_1}+\frac{\hat{\vec{p}}_1^2}{2m_n}-\mu+\zeta_n\right]F^+_{nn}(\vec{r}_1,t_1;\vec{r}_2,t_2)-i\sum_l V_{nl}F^+_{ll}(\vec{r}_1,t_1;\vec{r}_1,t_1)G_{nn}(\vec{r}_1,t_1;\vec{r}_2,t_2)=0$$

Let's define the order parameter $\Delta_n(\vec{r},t)$ as follows

$$\Delta_n(\vec{r},t)=i\sum_l V_{nl}F_{ll}(\vec{r},t;\vec{r},t),$$
$$\Delta^*_n(\vec{r},t)=-i\sum_l V_{nl}F^+_{ll}(\vec{r},t;\vec{r},t) \quad (1.10)$$

Then (1.8), (1.7) may be rewritten in the form

$$\left[i\frac{\partial}{\partial t_1}-\frac{\hat{\vec{p}}_1^2}{2m_n}+\mu-\zeta_n\right]G_{nn}(\vec{r}_1,t_1;\vec{r}_2,t_2)+\Delta_n(\vec{r}_1,t_1)F^+_{nn}(\vec{r}_1,t_1;\vec{r}_2,t_2)=$$
$$=\delta(\vec{r}_1-\vec{r}_2)\delta(t_1-t_2) \quad (1.11)$$

$$\left[i\frac{\partial}{\partial t_1}+\frac{\hat{\vec{p}}_1^2}{2m_n}-\mu+\zeta_n\right]F^+_{nn}(\vec{r}_1,t_1;\vec{r}_2,t_2)+\Delta^*_n(\vec{r}_1,t_1)G_{nn}(\vec{r}_1,t_1;\vec{r}_2,t_2)=0 \quad (1.12)$$

The pair of equations (1.11),(1.12) are Gor'kov equations for two-band superconductor.

The equation (1.10) is the self-consistency condition.

## 2. A time dependent Ginzburg-Landau equation

In the limit case $V_{nm} = 0$ we have $\Delta_n = 0$ and the system (1.11),(1.12) obtains the form

$$\left[i\frac{\partial}{\partial t_1} - \frac{\hat{\vec{p}}_1^2}{2m_n} + \mu - \zeta_n\right] G^{(0)}{}_{nn}(\vec{r}_1,t_1;\vec{r}_2,t_2) = \delta(\vec{r}_1 - \vec{r}_2)\delta(t_1 - t_2) \tag{2.1}$$

$$F^{+(0)}{}_{nn}(\vec{r}_1,t_1;\vec{r}_2,t_2) = 0 \tag{2.2}$$

The system (1.10), (1.11) we may rewrite as integral equations

$$G_{nn}(\vec{r}_1,t_1;\vec{r}_2,t_2) + \iint d\vec{r}_3 dt_3\, g^{(0)}{}_n(\vec{r}_1,t_1;\vec{r}_3,t_3)\Delta_n(\vec{r}_3,t_3)F^+_{nn}(\vec{r}_3,t_3;\vec{r}_2,t_2) =$$
$$= g^{(0)}{}_n(\vec{r}_1,t_1;\vec{r}_2,t_2) \tag{2.3}$$

$$F^+{}_{nn}(\vec{r}_1,t_1;\vec{r}_2,t_2) + \iint d\vec{r}_3 dt_3\, \tilde{g}^{(0)}_n(\vec{r}_1,t_1;\vec{r}_3,t_3)\Delta^*{}_n(\vec{r}_3,t_3)G_{nn}(\vec{r}_3,t_3;\vec{r}_2,t_2) = 0 \tag{2.4}$$

$$\left[i\frac{\partial}{\partial t_1} + \frac{\hat{\vec{p}}_1^2}{2m_n} - \mu + \zeta_n\right] \tilde{g}^{(0)}_n(\vec{r}_1,t_1;\vec{r}_2,t_2) = \delta(\vec{r}_1 - \vec{r}_2)\delta(t_1 - t_2)$$

$$\left[i\frac{\partial}{\partial t_1} - \frac{\hat{\vec{p}}_1^2}{2m_n} + \mu - \zeta_n\right] g^{(0)}{}_n(\vec{r}_1,t_1;\vec{r}_2,t_2) = \delta(\vec{r}_1 - \vec{r}_2)\delta(t_1 - t_2) \tag{2.5}$$

The functions $g^{(0)}_n(\vec{r}_1,t_1;\vec{r}_2,t_2)$ and $\tilde{g}^{(0)}_n(\vec{r}_1,t_1;\vec{r}_2,t_2)$ are functions of the differences $\vec{r}_1-\vec{r}_2, t_1-t_2$:

$$g^{(0)}_n(\vec{r}_1,t_1;\vec{r}_2,t_2) = g^{(0)}_n(\vec{r}_1-\vec{r}_2, t_1-t_2),$$
$$\tilde{g}^{(0)}_n(\vec{r}_1,t_1;\vec{r}_2,t_2) = \tilde{g}^{(0)}_n(\vec{r}_1-\vec{r}_2, t_1-t_2)$$

Then

$$g^{(0)}_n(\vec{r}_1,t_1;\vec{r}_2,t_2) = \iint g^{(0)}_n(\omega,\vec{p}) e^{i\vec{p}(\vec{r}_1-\vec{r}_2)-i\omega(t_1-t_2)} \frac{d\omega d^3 p}{(2\pi)^4}$$

$$[\omega - \varepsilon_n(\vec{p})] g^{(0)}_n(\omega,\vec{p}) = 1 \qquad (2.6)$$

$$\tilde{g}^{(0)}_n(\vec{r}_1,t_1;\vec{r}_2,t_2) = \iint \tilde{g}^{(0)}_n(\omega,\vec{p}) e^{i\vec{p}(\vec{r}_1-\vec{r}_2)-i\omega(t_1-t_2)} \frac{d\omega d^3 p}{(2\pi)^4},$$

$$[\omega + \varepsilon_n(\vec{p})] \tilde{g}^{(0)}_n(\omega,\vec{p}) = 1, \qquad (2.7)$$

$$\tilde{g}^{(0)}_n(\vec{r}_1,t_1;\vec{r}_2,t_2) = -g^{(0)}_n(\vec{r}_2,t_2;\vec{r}_1,t_1)$$

With (2.7) the equation (2.4) takes the form

$$F^+_{nn}(\vec{r}_1,t_1;\vec{r}_2,t_2) = -\iint d\vec{r}_3 dt_3 g^{(0)}_n(\vec{r}_3,t_3;\vec{r}_1,t_1) \Delta^*_n(\vec{r}_3,t_3) G_{nn}(\vec{r}_3,t_3;\vec{r}_2,t_2) \qquad (2.8)$$

To solve the system (2.6),(2.11) iteratively we write $\Delta_{nl}$ as $\lambda\Delta_{nl}$ and expand $G_{nm}(\vec{r}_1,t_1;\vec{r}_2,t_2)$, $F^+_{nm}(\vec{r}_1,t_1;\vec{r}_2,t_2)$ in the form

$$G_{nm}(\vec{r}_1,t_1;\vec{r}_2,t_2) = G^{(0)}_{nm}(\vec{r}_1,t_1;\vec{r}_2,t_2) + \lambda^2 G^{(2)}_{nm}(\vec{r}_1,t_1;\vec{r}_2,t_2) + ...$$

$$F^+_{nm}(\vec{r}_1,t_1;\vec{r}_2,t_2) = \lambda F^{+(1)}_{nm}(\vec{r}_1,t_1;\vec{r}_2,t_2) + \lambda^3 F^{+(3)}_{nm}(\vec{r}_1,t_1;\vec{r}_2,t_2)...$$

Substituting these forms into (1.21),(1.22) and associating equal powers of $\lambda$ we have

$$F^{+(1)}{}_{nn}(\vec{r}_1,t_1;\vec{r}_2,t_2) = -\iint d\vec{r}_3 dt_3 g^{(0)}{}_n(\vec{r}_3,t_3;\vec{r}_1,t_1)\Delta^*{}_n(\vec{r}_3,t_3)g^{(0)}{}_n(\vec{r}_3,t_3;\vec{r}_2,t_2) \qquad (2.9)$$

$$G^{(2)}{}_{nm}(\vec{r}_1,t_1;\vec{r}_2,t_2) = -\iint d\vec{r}_3 dt_3 g^{(0)}{}_n(\vec{r}_1,t_1;\vec{r}_3,t_3)\Delta_n(\vec{r}_3,t_3)F^{+(1)}_{nn}(\vec{r}_3,t_3;\vec{r}_2,t_2) =$$
$$= \iint d\vec{r}_3 dt_3 g^{(0)}{}_n(\vec{r}_1,t_1;\vec{r}_3,t_3)\Delta_n(\vec{r}_3,t_3)\iint d\vec{r}_4 dt_4 g^{(0)}{}_n(\vec{r}_4,t_4;\vec{r}_3,t_3)\Delta^*{}_n(\vec{r}_4,t_4)g^{(0)}{}_n(\vec{r}_4,t_4;\vec{r}_2,t_2) = \qquad (2.10)$$
$$= \iint d\vec{r}_3 dt_3 \iint d\vec{r}_4 dt_4 g^{(0)}{}_n(\vec{r}_1,t_1;\vec{r}_3,t_3)\Delta_n(\vec{r}_3,t_3)g^{(0)}{}_n(\vec{r}_4,t_4;\vec{r}_3,t_3)\Delta^*{}_n(\vec{r}_4,t_4)g^{(0)}{}_n(\vec{r}_4,t_4;\vec{r}_2,t_2)$$

$$F^{+(3)}{}_{nn}(\vec{r}_1,t_1;\vec{r}_2,t_2) = \iint d\vec{r}_3 dt_3 g^{(0)}{}_n(\vec{r}_3,t_3;\vec{r}_1,t_1)\Delta^*{}_n(\vec{r}_3,t_3)G^{(2)}{}_{nn}(\vec{r}_3,t_3;\vec{r}_2,t_2) =$$
$$= \iint d\vec{r}_3 dt_3 \iint d\vec{r}_4 dt_4 \iint d\vec{r}_5 dt_5 g^{(0)}{}_n(\vec{r}_3,t_3;\vec{r}_1,t_1)\Delta^*{}_n(\vec{r}_3,t_3)g^{(0)}{}_n(\vec{r}_3,t_3;\vec{r}_4,t_4)\times \qquad (2.11)$$
$$\times \Delta_n(\vec{r}_4,t_4)g^{(0)}{}_n(\vec{r}_5,t_5;\vec{r}_4,t_4)\Delta^*{}_n(\vec{r}_5,t_5)g^{(0)}{}_n(\vec{r}_5,t_5;\vec{r}_2,t_2)$$

Adding up the results (1.10), (2.9), (2.11) we obtain

$$\Delta^*{}_n(\vec{r},t) = -i\sum_m V_{nm}F^+{}_{mm}(\vec{r},t;\vec{r},t) = -i\sum_m V_{nm}\left[F^{+(1)}{}_{mm}(\vec{r},t;\vec{r},t) + F^{+(3)}{}_{mm}(\vec{r},t;\vec{r},t)\right] =$$
$$= -i\sum_m V_{nm}\iint d\vec{r}_1 dt_1 g^{(0)}{}_m(\vec{r}_1,t_1;\vec{r},t)\Delta^*{}_m(\vec{r}_1,t_1)g^{(0)}{}_m(\vec{r}_1,t_1;\vec{r},t) +$$
$$+ i\sum_m V_{nm}\iint d\vec{r}_1 dt_1 \iint d\vec{r}_2 dt_2 \iint d\vec{r}_3 dt_3 g^{(0)}{}_m(\vec{r}_1,t_1;\vec{r},t)\Delta^*{}_m(\vec{r}_1,t_1)g^{(0)}{}_m(\vec{r}_1,t_1;\vec{r}_2,t_2) \times \qquad (2.12)$$
$$\times \Delta_m(\vec{r}_2,t_2)g^{(0)}{}_m(\vec{r}_3,t_3;\vec{r}_2,t_2)\Delta^*{}_m(\vec{r}_3,t_3)g^{(0)}{}_m(\vec{r}_3,t_3;\vec{r},t)$$

Let us denote $W_{mn}$ elements of inverse matrix for $(V_{nm})$ then (2.12) obtains the form

$$\sum_n W_{mn} \Delta^*_n(\vec{r},t) = -i \iint d\vec{r}_1 dt_1 g^{(0)}_m(\vec{r}_1,t_1;\vec{r},t) \Delta^*_m(\vec{r}_1,t_1) g^{(0)}_m(\vec{r}_1,t_1;\vec{r},t) +$$
$$+ i \iint d\vec{r}_1 dt_1 \iint d\vec{r}_2 dt_2 \iint d\vec{r}_3 dt_3 g^{(0)}_m(\vec{r}_1,t_1;\vec{r},t) \Delta^*_m(\vec{r}_1,t_1) g^{(0)}_m(\vec{r}_1,t_1;\vec{r}_2,t_2) \times \qquad (2.13)$$
$$\times \Delta_m(\vec{r}_2,t_2) g^{(0)}_m(\vec{r}_3,t_3;\vec{r}_2,t_2) \Delta^*_m(\vec{r}_3,t_3) g^{(0)}_m(\vec{r}_3,t_3;\vec{r},t)$$

We now expand $\Delta_m(\vec{r}_1,t_1)$ in the first integral of the r.h.s. of eq. (2.13) as follows

$$\Delta_m(\vec{r}_1,t_1) = \Delta_m(\vec{r},t) + \frac{\partial \Delta_m(\vec{r},t)}{\partial x_\alpha}(x_{1\alpha}-x_\alpha) + \frac{1}{2}\frac{\partial^2 \Delta_m(\vec{r},t)}{\partial x_\alpha \partial x_\beta}(x_{1\alpha}-x_\alpha)(x_{1\beta}-x_\beta) + $$
$$+ \frac{\partial \Delta_m(\vec{r},t)}{\partial t}(t_1-t) \qquad (2.14)$$

where $(x_1, x_2, x_3) = \vec{r}$.

In the second integral we take the $\Delta_m(\vec{r}_1,t_1), \Delta_m(\vec{r}_2,t_2) \Delta_m(\vec{r}_3,t_3)$ just equal to $\Delta_m(\vec{r},t)$.

Using (2.14), we put eq. (2.13) in the following form

$$\sum_n W_{mn} \Delta_n(\vec{r},t) = A_m \Delta_m(\vec{r},t) - \Gamma_m \frac{\partial \Delta_m(\vec{r},t)}{\partial t} + D^{\alpha\beta}_m \frac{\partial^2 \Delta_m(\vec{r},t)}{\partial x_\alpha \partial x_\beta} - B_m \Delta_m(\vec{r},t)|\Delta_m(\vec{r},t)|^2 \qquad (2.15)$$
$$(m=1,2)$$

Where

$$A_m = -i \iint d\vec{r}_1 dt_1 \left[ g^{(0)}_m(\vec{r}_1,t_1;\vec{r},t) \right]^2 \qquad (2.16)$$

$$\Gamma_m = -i \iint d\vec{r}_1 dt_1 (t_1-t) \left[ g^{(0)}_m(\vec{r}_1,t_1;\vec{r},t) \right]^2 \qquad (2.17)$$

$$D^{\alpha\beta}_m = -\frac{i}{2} \iint d\vec{r}_1 dt_1 (x_{1\alpha}-x_\alpha)(x_{1\beta}-x_\beta) \left[ g^{(0)}_m(\vec{r}_1,t_1;\vec{r},t) \right]^2 \qquad (2.18)$$

$$B_m = -i \iint d\vec{r}_1 dt_1 \iint d\vec{r}_2 dt_2 \iint d\vec{r}_3 dt_3 g^{(0)}{}_m(\vec{r}_1,t_1;\vec{r},t)g^{(0)}{}_m(\vec{r}_1,t_1;\vec{r}_2,t_2) \times$$
$$\times g^{(0)}{}_m(\vec{r}_3,t_3;\vec{r}_2,t_2)g^{(0)}{}_m(\vec{r}_3,t_3;\vec{r},t) \tag{2.19}$$

The eq.(2.15) is the Ginzburg-Landau equation for two-band superconductor

$$\Gamma_m \frac{\partial \Delta_m(\vec{r},t)}{\partial t} = \sum_{n=1,2}[A_m \delta_{nm} - W_{mn}]\Delta_n(\vec{r},t) + D_m^{\alpha\beta}\frac{\partial^2 \Delta_m(\vec{r},t)}{\partial x_\alpha \partial x_\beta} - B_m \Delta_m(\vec{r},t)|\Delta_m(\vec{r},t)|^2 \tag{2.20}$$
$(m=1,2)$

or

$$\Gamma_1 \frac{\partial \Delta_1(\vec{r},t)}{\partial t} = (A_1 - W_{11})\Delta_1(\vec{r},t) - W_{12}\Delta_2(\vec{r},t) + D_1^{\alpha\beta}\frac{\partial^2 \Delta_1(\vec{r},t)}{\partial x_\alpha \partial x_\beta} - B_1 \Delta_1(\vec{r},t)|\Delta_1(\vec{r},t)|^2 \tag{2.21}$$

$$\Gamma_2 \frac{\partial \Delta_2(\vec{r},t)}{\partial t} = -W_{21}\Delta_1(\vec{r},t) + (A_2 - W_{22})\Delta_2(\vec{r},t) + D_2^{\alpha\beta}\frac{\partial^2 \Delta_2(\vec{r},t)}{\partial x_\alpha \partial x_\beta} - B_2 \Delta_2(\vec{r},t)|\Delta_2(\vec{r},t)|^2 \tag{2.22}$$

Equations (2.21), (2.22) can be rewritten as

$$\Gamma_1 \frac{\partial \Delta_1}{\partial t} = -\frac{\delta F}{\delta \Delta_1^*},$$
$$\Gamma_2 \frac{\partial \Delta_2}{\partial t} = -\frac{\delta F}{\delta \Delta_2^*}$$

where

$$F = \int d\vec{r} \left[ \frac{1}{2}(W_{11} - A_1)|\Delta_1|^2 + \frac{1}{2}(W_{22} - A_2)|\Delta_2|^2 + W_{12}\Delta_1^*\Delta_2 + W_{21}\Delta_2^*\Delta_1 + \right.$$
$$\left. + \frac{1}{2} D_1^{\alpha\beta} \frac{\partial \Delta_1}{\partial x_\alpha} \frac{\partial \Delta^*_1}{\partial x_\beta} + \frac{1}{2} D_2^{\alpha\beta} \frac{\partial \Delta_2}{\partial x_\alpha} \frac{\partial \Delta^*_2}{\partial x_\beta} + \frac{1}{4} B_1 |\Delta_1|^4 + \frac{1}{4} B_2 |\Delta_2|^4 \right]$$

is the GL functional.

## 3. The coefficients of the model.

At present section we derive the formulas for evaluation of the coefficients $A_m, \Gamma_m, D_m^{\alpha\beta}, B_m$. Since the functions $g^{(0)}_n(\vec{r}_1, t_1; \vec{r}_2, t_2)$ are functions of the differences $\vec{r}_1 - \vec{r}_2, t_1 - t_2$:

$$g_n^{(0)}(\vec{r}_1, t_1; \vec{r}_2, t_2) = g_n^{(0)}(\vec{r}_1 - \vec{r}_2, t_1 - t_2),$$

We obtain

$$A_m = -i \iint d\vec{r}_1 dt_1 \left[ g_m^{(0)}(\vec{r}_1, t_1; \vec{r}, t) \right]^2 = -i \iint d\vec{r}_1 dt_1 \left[ g_m^{(0)}(\vec{r}_1 - \vec{r}, t_1 - t) \right]^2 =$$
$$= -i \int d\vec{r}_1 \int_{-\infty}^{t} dt_1 \left[ g_m^{(0)}(\vec{r}_1 - \vec{r}, t_1 - t) \right]^2 - i \int d\vec{r}_1 \int_{t}^{\infty} dt_1 \left[ g_m^{(0)}(\vec{r}_1 - \vec{r}, t_1 - t) \right]^2 =$$
$$= -i \int d\vec{r}_1 \int_{-\infty}^{0} d\tau \left[ g_m^{<}(\vec{r}_1 - \vec{r}, \tau) \right]^2 - i \int d\vec{r}_1 \int_{0}^{\infty} d\tau \left[ g_m^{>}(\vec{r}_1 - \vec{r}, \tau) \right]^2 =$$
$$= -i \int d\vec{r} \int_{-\infty}^{0} d\tau \left[ g_m^{<}(\vec{r}, \tau) \right]^2 - i \int d\vec{r} \int_{0}^{\infty} d\tau \left[ g_m^{>}(\vec{r}, \tau) \right]^2 \quad (3.1)$$

where $g_m^{<}(\vec{r}, t)$, $g_m^{>}(\vec{r}, t)$ are the correlation functions which Fourier transform is

$$g_m^{>}(\vec{p}, t) = -i \frac{e^{-i\varepsilon_m(\vec{p})t}}{e^{-\beta\varepsilon_m(\vec{p})} + 1},$$
$$g_m^{<}(\vec{p}, t) = i \frac{e^{-i\varepsilon_m(\vec{p})t}}{e^{\beta\varepsilon_m(\vec{p})} + 1}, \quad (3.2)$$
$$\beta = \frac{1}{k_B T}$$

and

$$\begin{cases} f(\vec{r}) = \int f(\vec{p}) e^{i\vec{p}\vec{r}} \dfrac{d\vec{p}}{(2\pi)^3} \\ f(\vec{p}) = \int f(\vec{r}) e^{-i\vec{p}\vec{r}} d\vec{r} \end{cases}$$

Then

$$A_m = \frac{1}{2(2\pi)^3} \int \frac{\tanh(\frac{1}{2}\beta\varepsilon_m(\vec{p}))}{\varepsilon_m(\vec{p})} d\vec{p} =$$

$$= \frac{\hbar^3}{2} \int_0^{\hbar\omega_D} \frac{\tanh(\frac{1}{2}\beta\varepsilon_m)}{\varepsilon_m} \cdot N_m(\varepsilon) d\varepsilon_m = \qquad (3.3)$$

$$= \frac{\hbar^3 N_m(0)}{2} \int_0^{\hbar\omega_D} \frac{d\varepsilon}{\varepsilon} \tanh(\frac{1}{2}\beta\varepsilon) = \frac{\hbar^3 N_m(0)}{2} \int_0^{\frac{\beta\hbar\omega_D}{2}} \frac{\tanh x}{x} dx$$

where $N_m(0)$ is the density of states at the Fermi surface of $m$-tm band, $\omega_D$ is the Debye frequency. For most superconductors $\dfrac{\hbar\omega_D}{2k_B T_c} \gg 1$, where $T_c$ is the critical temperature.

The integral (3.3) can be evaluated using

$$\int_0^a \frac{\tanh x}{x} dx = \tanh a \cdot \ln a - \int_0^a \frac{\ln x}{\cosh^2 x} dx$$

For $a \gg 1$ we obtain

$$\int_0^a \frac{\tanh x}{x} dx \approx \ln a - \int_0^\infty \frac{\ln x}{\cosh^2 x} dx = \ln a - \ln\left(\frac{\pi}{4e^C}\right) = \ln\left(\frac{4e^C a}{\pi}\right)$$

with $C \approx 0.577$ being the Euler constant[43].

Therefore

14$$A_m(T) = \frac{N_m(0)\hbar^3}{4\pi} \ln\left(\frac{2e^C \hbar \omega_D}{\pi k_B T}\right) \quad (3.4)$$

The critical temperature of a two-band superconductor is determined by the condition[15,17,18,19,20,21,27,33]

$$\det\begin{pmatrix} A_1(T_c) - W_{11} & -W_{12} \\ -W_{21} & A_2(T_c) - W_{22} \end{pmatrix} = 0 \quad (3.5)$$

The damping constants are

$$\Gamma_m = -i\iint d\vec{r_1} dt_1 (t_1-t)\left[g_m^{(0)}(\vec{r_1},t_1;\vec{r},t)\right]^2 =$$

$$= -i\iint d\vec{r_1} dt_1 (t_1-t)\left[g_m^{(0)}(\vec{r_1}-\vec{r},t_1-t)\right]^2 =$$

$$= -i\int d\vec{r_1} \int_{-\infty}^{t} dt_1 (t_1-t)\left[g_m^{(0)}(\vec{r_1}-\vec{r},t_1-t)\right]^2 -i\int d\vec{r_1} \int_{t}^{\infty} dt_1 (t_1-t)\left[g_m^{(0)}(\vec{r_1}-\vec{r},t_1-t)\right]^2 =$$

$$= -i\int d\vec{r_1} \int_{-\infty}^{0} d\tau \tau \left[g_m^{<}(\vec{r_1}-\vec{r},\tau)\right]^2 -i\int d\vec{r_1} \int_{0}^{\infty} d\tau \tau \left[g_m^{>}(\vec{r_1}-\vec{r},\tau)\right]^2 =$$

$$= -i\int d\vec{r} \int_{-\infty}^{0} d\tau \tau \left[g_m^{<}(\vec{r},\tau)\right]^2 -i\int d\vec{r} \int_{0}^{\infty} d\tau \tau \left[g_m^{>}(\vec{r},\tau)\right]^2 =$$

$$= -i\int \frac{d\vec{p}}{(2\pi)^3} \int_{-\infty}^{0} d\tau \tau \left[g_m^{<}(\vec{p},\tau)\right]^2 -i\int \frac{d\vec{p}}{(2\pi)^3} \int_{0}^{\infty} d\tau \tau \left[g_m^{>}(\vec{p},\tau)\right]^2$$

Using (3.2) we obtain

$$\Gamma_m = -\frac{i}{4(2\pi)^3} \int \frac{\tanh\left(\frac{1}{2}\beta\varepsilon_m(\vec{p})\right)}{(\varepsilon_m(\vec{p})+i0)^2} d\vec{p} = -\frac{i\hbar^3}{4}\int N_m(\varepsilon)\frac{\tanh\left(\frac{1}{2}\beta\varepsilon_m\right)}{(\varepsilon_m+i0)^2} d\varepsilon_m =$$

$$= -\frac{i\hbar^3 \beta N_m(0)}{8}\int \frac{\tanh x}{(x+i0)^2} dx = \frac{\pi \hbar^3 N_m(0)}{16 k_B T_c} \quad (3.6)$$

The diffusion constants are

$$D_n^{\alpha\beta} = -\frac{i}{2}\iint d\vec{r}\,dt\, x_\alpha x_\beta \left[g_n^{(0)}(\vec{r},t)\right]^2 = -\frac{\beta}{4(2\pi)^3}\int d\vec{p}\,\frac{\partial^2}{\partial p_\alpha \partial p_\beta}\left[\frac{\tanh(\beta\varepsilon_n(\vec{p}))}{\beta\varepsilon_n(\vec{p})}\right]$$

The integral can be evaluated using

$$\frac{\tanh x}{x} = \sum_{n=-\infty}^{\infty}\frac{1}{x^2+\left[\pi\left(n+\frac{1}{2}\right)\right]^2}$$

After the substitution we find

$$D_n^{\alpha\beta} = -\frac{\beta}{4(2\pi)^3}\sum_{n=-\infty}^{\infty}\int d\vec{p}\,\frac{\partial^2}{\partial p_\alpha \partial p_\beta}\left[\frac{1}{\beta^2\varepsilon_n^2(\vec{p})+\pi^2\left(n+\frac{1}{2}\right)^2}\right] \approx$$

$$\approx \frac{\beta^3}{2(2\pi)^3}\sum_{n=-\infty}^{\infty}\int d\vec{p}\,\frac{\frac{\partial\varepsilon_n(\vec{p})}{\partial p_\alpha}\cdot\frac{\partial\varepsilon_n(\vec{p})}{\partial p_\alpha}}{\left[\beta^2\varepsilon_n^2(\vec{p})+\pi^2\left(n+\frac{1}{2}\right)^2\right]^2} \approx \frac{\beta^3}{2(2\pi)^3}\sum_{n=-\infty}^{\infty}\int d\vec{p}\,\frac{v_{F\alpha}\cdot v_{F\beta}}{\left[\beta^2\varepsilon_n^2(\vec{p})+\pi^2\left(n+\frac{1}{2}\right)^2\right]^2} =$$

$$= \frac{\beta^3 v_F^2 \delta_{\alpha\beta}}{6(2\pi)^3}\sum_{n=-\infty}^{\infty}\int d\vec{p}\,\frac{1}{\left[\beta^2\varepsilon_n^2(\vec{p})+\pi^2\left(n+\frac{1}{2}\right)^2\right]^2}$$

where $\vec{v}_F = \dfrac{\vec{p}_F}{m_n}$ is the Fermi velocity. Inserting $d\vec{p}=(2\pi\hbar)^3 N_n(\varepsilon)d\varepsilon$ [44] we obtain

$$D_n^{\alpha\beta} = \frac{\beta^3 v_F^2 \delta_{\alpha\beta} N_n(0)\hbar^3}{6} \sum_{n=-\infty}^{\infty} \int_0^{\infty} \frac{d\varepsilon}{\left[\beta^2\varepsilon^2 + \pi^2\left(n+\frac{1}{2}\right)^2\right]^2} =$$

$$= \frac{\beta^2 v_F^2 \delta_{\alpha\beta} N_n(0)\hbar^3}{6} \sum_{n=-\infty}^{\infty} \int_0^{\infty} \frac{dx}{\left[x^2 + \pi^2\left(n+\frac{1}{2}\right)^2\right]^2} = \frac{\beta^2 v_F^2 \delta_{\alpha\beta} N_n(0)\hbar^3}{6} \sum_{n=-\infty}^{\infty} \frac{\pi}{4\left(n+\frac{1}{2}\right)^3} =$$

$$= \frac{7\pi\beta^2 v_F^2 \delta_{\alpha\beta} N_n(0)\hbar^3}{24} \varsigma(3) \equiv D_n \delta_{\alpha\beta}$$

Here we use

$$\sum_{n=0}^{\infty} \frac{1}{\left(n+\frac{1}{2}\right)^z} = (2^z - 1)\varsigma(z)$$

Where $\sum_{n=1}^{\infty} \frac{1}{n^z} = \varsigma(z)$ is the Riemann $\varsigma$ function.

The Green's function satisfies the anti-periodic boundary condition
$g_m^{(0)}(\vec{r}_1,t_1;\vec{r},t)\big|_{t_1=0} = -g_m^{(0)}(\vec{r}_1,t_1;\vec{r},t)\big|_{t_1=-i\beta}$ we express $g_m^{(0)}$ as Fourier series [45]

$$g_m^{(0)}(\vec{p},t-t') = \frac{i}{\beta} \sum_v \frac{e^{-iz_v(t-t')}}{z_v - \varepsilon_m(\vec{p})},$$

$$z_v = \frac{\pi i v}{\beta}$$

(3.7)

where the sum is taken to run over all odd integers $(v = 2n-1, n \in Z)$. If we then substitute (3.7) into the (2.19) we obtain

$$B_m = \frac{1}{\pi\beta} \sum_v \int d\vec{p} \frac{1}{\left[\varepsilon_m^2(\vec{p}) + \frac{\pi^2 v^2}{\beta^2}\right]^2} = \frac{8\pi^2 N_m(0)\hbar^3}{\beta} \sum_v \int_0^{\infty} \frac{d\varepsilon}{\left(\varepsilon^2 + \frac{\pi^2 v^2}{\beta^2}\right)^2} =$$

$$= 2N_m(0)\hbar^3 \beta^2 \sum_{n \in Z} \frac{1}{(2n-1)^3} = \frac{7N_m(0)\hbar^3 \varsigma(3)}{4(k_B T)^2}$$

## 4. The linearized model.

The equations (2.21), (2.22) are nonlinear differential equations of the diffusion type to describe the space-time variation of the order parameters near the critical temperature. The self consistent superconducting interaction drives the order parameters towards its equilibrium values $\bar{\Delta}_1, \bar{\Delta}_2$ which are the static uniform solution

$$0 = (A_1 - W_{11})\bar{\Delta}_1 - W_{12}\bar{\Delta}_2 - B_1\bar{\Delta}_1^3$$

$$0 = -W_{21}\bar{\Delta}_1 + (A_2 - W_{22})\bar{\Delta}_2 - B_2\bar{\Delta}_2^3$$

For $T > T_c$ we have $\bar{\Delta}_1 = \bar{\Delta}_2 = 0$ (symmetrical phase) while for $T < T_c$ we have $\bar{\Delta}_1, \bar{\Delta}_2 \propto \sqrt{\tau}$ (unsymmetrical phase) where $\tau = 1 - \dfrac{T}{T_c}$. In this section we linearize our model about equilibrium: $\Delta_n(\vec{r},t) = \bar{\Delta}_n + \eta_n(\vec{r},t)$ where $\eta_n(\vec{r},t)$ is the deviation of the order parameters from its real equilibrium value. In the linear approximation $\eta_n(\vec{r},t)$ satisfy the set of equations

$$\Gamma_1 \frac{\partial \eta_1}{\partial t} = \left(A_1 - W_{11} - 3B_1\bar{\Delta}_1^2\right)\eta_1 - W_{12}\eta_2 + D_1\Pi^2\eta_1$$
$$\Gamma_2 \frac{\partial \eta_2}{\partial t} = -W_{21}\eta_1 + \left(A_2 - W_{22} - 3B_2\bar{\Delta}_2^2\right)\eta_2 + D_2\Pi^2\eta_2$$
(4.1)

where $\Pi^2$ is the Laplace operator. In a spatially homogeneous system the eq.(4.1) takes the form

$$\frac{d\vec{\eta}(t)}{dt} = -L(\tau)\vec{\eta}(t) \qquad (4.2)$$

where $\vec{\eta}(t) = \begin{pmatrix} \eta_1(t) \\ \eta_2(t) \end{pmatrix}$, $L(\tau) = \begin{pmatrix} \dfrac{3B_1\bar{\Delta}_1^2 + W_{11} - A_1}{\Gamma_1} & \dfrac{W_{12}}{\Gamma_1} \\ \dfrac{W_{21}}{\Gamma_2} & \dfrac{3B_2\bar{\Delta}_2^2 + W_{22} - A_2}{\Gamma_2} \end{pmatrix}$. The general solution of eq. (4.2) is

$$\vec{\eta}(t) = C_1\vec{u}^{(1)}e^{\lambda_1(\tau)t} + C_2\vec{u}^{(2)}e^{\lambda_2(\tau)t} \qquad (4.3)$$

where $C_1, C_2$ are arbitrary constants, $\lambda_1(\tau), \lambda_2(\tau)$ are eigenvalues of $L(\tau)$ and $\vec{u}^{(1)}, \vec{u}^{(2)}$ are the corresponding eigenvectors. The system has two relaxation times: $\vartheta_1 = -\frac{1}{\lambda_1}, \vartheta_2 = -\frac{1}{\lambda_2}, (\vartheta_1 > \vartheta_2)$. The eq. (3.5) leads to

$$\lim_{\tau \to 0} \lambda_1(\tau) = 0$$
$$\lim_{\tau \to 0} \lambda_2(\tau) = -\mu < 0$$

Thus, for $t \gg \vartheta_2 = \mu^{-1}$ the second term in (4.3) is negligible. We have

$$\vec{\eta}(t) \underset{t \gg \mu^{-1}}{\cong} C_1 \vec{u}_1 e^{\lambda_1(\tau)t} \tag{4.4}$$

where $\vec{u}^{(1)} = \begin{pmatrix} u_1^{(1)} \\ u_2^{(1)} \end{pmatrix}$ is a constant vector. It is readily checked that

$$\frac{\eta_1(t)}{\eta_2(t)} \underset{t \gg \mu^{-1}}{=} \frac{u_1^{(1)}}{u_2^{(1)}} = const$$

So the ratio $\frac{\eta_1(t)}{\eta_2(t)}$ comes out to be time independent. Order parameters of a two-band superconductor are strongly correlated in the limit $t \to \infty$. Both order parameters vary on the same time scale. This conclusion is in agreement with [33]. The motion of pairs in one band causes a change in another one. The first subsystem follows adiabatically the second. We shall call this regime the "adiabatic regime".

If $\lambda_1(\tau)$ is expanded in powers of $\tau$, within the GL accuracy we can put $\lambda_1(\tau) = const \cdot \tau$. The largest relaxation time of the system is $\vartheta_1 = -\frac{1}{\lambda_1(\tau)} \propto \frac{1}{\tau}$. Thus, in this approximation $\frac{\partial}{\partial t} \propto \tau, and \frac{\partial \Delta_1}{\partial t} \propto \tau^{3/2}, \frac{\partial \Delta_2}{\partial t} \propto \tau^{3/2}$.

## 5. The nonlinear model in adiabatic regime.

Our goal is to obtain the set of decoupled equations for $\Delta_1, \Delta_2$. We now recall [33] that in the one-band GL equation

$$\gamma \frac{\partial \Delta}{\partial t} = a\Delta - b\Delta |\Delta|^2 + D\Pi^2 \Delta$$

all terms are of the same order with respect to $\tau: a \propto \tau^1, b \propto \tau^0, \Delta \propto \tau^{1/2}, \partial/\partial t \propto \tau^1, \Pi^2 \propto \xi^{-2} \propto \tau$ where $\xi$ is the coherence length.

Following [33] we express $\Delta_2$ from eq.(2.21) and substitute the result in eq. (2.22), keeping only terms up to order $\tau^{3/2}$:

$$\gamma \frac{\partial \Delta_1}{\partial t} = a\Delta_1 - b_1 \Delta_1 |\Delta_1|^2 + K\Pi^2 \Delta_1 \qquad (5.1)$$

with

$$\gamma = \Gamma_1(A_2 - W_{22}) + \Gamma_2(A_1 - W_{11}), \quad a = (A_1 - W_{11})(A_2 - W_{22}) - W_{12}W_{21},$$
$$K = D_1(A_2 - W_{22}) + D_2(A_1 - W_{11}), \quad b_1 = B_1(A_2 - W_{22}) + \frac{B_2(A_1 - W_{11})^3}{W_{12}^2}$$

The term with $\frac{\partial^2 \Delta_1}{\partial t^2}$ is neglected.

Similarly, one obtains an equation for $\Delta_2$:

$$\gamma \frac{\partial \Delta_2}{\partial t} = a\Delta_2 - b_2 \Delta_2 |\Delta_2|^2 + K\Pi^2 \Delta_2 \qquad (5.2)$$

with $b_2 = B_2(A_1 - W_{11}) + \frac{B_1(A_2 - W_{22})^3}{W_{21}^2}$. Within the GL accuracy $a = k\tau, k > 0$.

We obtain upon dividing by $a$ of both equations (5.1),(5.2):

$$\vartheta \frac{\partial \Delta_1}{\partial t} = \Delta_1(1 - \frac{|\Delta_1|^2}{|\bar{\Delta}_1|^2}) + \xi^2 \Pi^2 \Delta_1,$$
$$\vartheta \frac{\partial \Delta_2}{\partial t} = \Delta_2(1 - \frac{|\Delta_2|^2}{|\bar{\Delta}_2|^2}) + \xi^2 \Pi^2 \Delta_2 \qquad (5.3)$$

where $\vartheta = \dfrac{\gamma}{a} = \dfrac{\gamma}{k\tau}$ is the largest relaxation time of the system,

$\xi = \sqrt{\dfrac{K}{a}} = \sqrt{\dfrac{K}{k\tau}}$, $\bar{\Delta}_1 = \sqrt{\dfrac{a}{b_1}}, \bar{\Delta}_2 = \sqrt{\dfrac{a}{b_2}}$.

The ratio $\dfrac{\Delta_1}{\bar{\Delta}_1} = \dfrac{\Delta_2}{\bar{\Delta}_2} = \Psi$ satisfies the equation

$$\vartheta \frac{\partial \Psi}{\partial t} = \Psi(1-|\Psi|^2) + \xi^2 \Pi^2 \Psi \qquad (5.4)$$

The two TDGL equations for the two-band case are reduced to a single equation for the normalized order parameter $\Psi(\vec{r},t)$.

## Acknowledgments

We are indebted to  J.Berger, V. Zhuravlev for useful discussions and help.